\documentclass[11pt,twoside]{article}

\usepackage{asp2006}
\usepackage{epsf}

\begin{document}
\title{Structural propeties of disk galaxies: \\ 
       The intrinsic equatorial ellipticity of bulges}   
\author{J. M\'endez-Abreu}
\affil{INAF-Osservatorio Astronomico di Padova, Padova, Italy}
\author{\rm J. A. L. Aguerri} 
\affil{Instituto de Astrof\'isica de Canarias, La Laguna, Spain}
\author{\rm E. M. Corsini} 
\affil{Dipartimento di Astronomia, Universit\`a di Padova, Padova, Italy}
\author{\rm E. Simonneau} 
\affil{Institute d'Astrophysique de Paris, C.N.R.S., Paris, France }    

\begin{abstract} 

The structural parameters of a magnitude-limited sample of 148
unbarred S0-Sb galaxies were derived to study the correlations between
bulge and disk parameters as well as the probability distribution
function (PDF) of the intrinsic equatorial ellipticity of bulges. A
new algorithm (GASP2D) was used to perform the bidimensional
bulge-disk decomposition of the $J$-band galaxy images extracted from
the archive of the 2MASS survey. The PDF of intrinsic ellipticities
was derived from the distribution of the observed ellipticities of the
bulges and misalignments between the the bulges and disks. About
$80\%$ of the observed bulges are not oblate but triaxial
ellipsoids. Their mean axial ratio in the equatorial plane is $\langle
B/A \rangle=0.85$. There is not significant dependence of their PDF on
morphology, light concentration or luminosity. This has to be
explained by the different scenarios of bulge formation.

\end{abstract}

\section{Structural parameters of the bulges and disks}

The structural parameters of the bulges and disks of a
magnitude-limited sample of 148 S0-Sb galaxies were derived from 2MASS
$J-$band images. We used GASP2D, our new code for two-dimensional
photometric decomposition (M\'endez-Abreu et al. 2008).
The S\'ersic bulge and exponential disk were assumed to be
characterized by elliptical and concentric isophotes with constant
(but possibly different) ellipticity and position angle.
We found that the surface-brightness radial profiles of larger bulges
are more centrally peaked than those of the smaller bulges. Larger
bulges have a lower effective surface brightness and higher absolute
luminosities. They reside in larger disks, as revealed by the
correlation between central velocity dispersion and disk scalelenght.
This reveals a strong coupling between bulges and disks. Larger disks
have a lower central surface brightness.

\section{The equatorial intrinsic ellipticity of bulges}

The determination of the intrinsic shape of bulges is an ill-posed
problem, that can be tackled only in a statistical way. A new approach
to derive the PDF of intrinsic equatorial ellipticities using only
photometric data was used (M\'endez-Abreu et al. 2008). Fig.~1 shows
the PDF obtained from our galaxy sample. The significant decrease of
probability for $E<0.07$ (or equivalently $B/A<0.93$) suggests that
the shape of bulges is elliptical rather than circular.
The average ellipticity $\langle E \rangle = 0.16$ ($\langle B/A
\rangle = 0.85$) is in agreement with previous findings by Bertola et
al. (1991) and Fathi \& Peletier (2003). The
$\rm{PDF}(E)$ is not related to morphology, light concentration, or
luminosity of bulges.

\begin{figure}
  \plotone{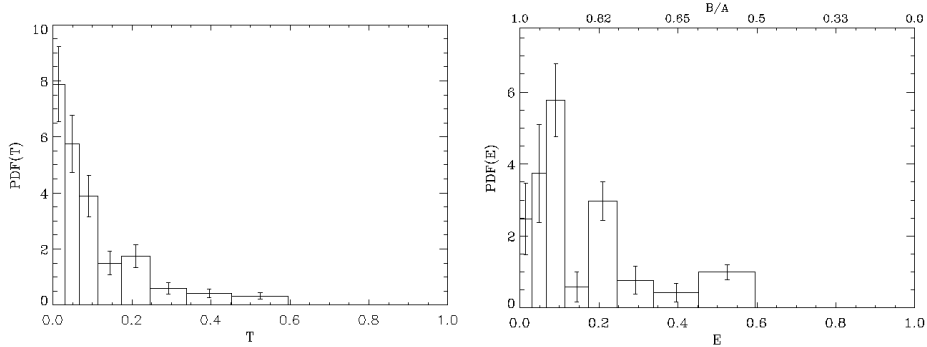}
  \caption{Left panel: \rm{PDF} of our observable $T$ which is
    calculated from the twist angle ($\delta$) between the bulge and
    disk, bulge ellipticity ($\epsilon$) and galaxy inclination
    ($\theta$). Right panel: \rm{PDF} of $E$ obtained from the
    inversion of $T$. $E$ is related to the intrinsic axis ratio by
    $E$=$(1-B/A)/(1+B/A)$ where $A$ and $B$ are the semi-axis of the
    bulge ellipsoid. The probability is normalized over 10 bins
    geometrically distributed, the width of the first bin is 0.03 and
    the width ratio is 1.25. Error bars correspond to a Poisson
    statistics.}
\end{figure}

\section{Discussion and conclusions}

The photometric and kinematic parameters of the bulges of our sample
galaxies follow the same fundamental and photometric planes of
elliptical galaxies, supporting the idea that both are formed in the
same way. However, the correlations between the bulge and disk
structural parameters are usually interpreted as an indication of
secular evolution processes. Therefore, the above relations are not
enough to distinguish between bulges formed by early dissipative
collapse, merging or secular evolution. All these scenarios could be
tested against the PDF of the intrinsic equatorial ellipticity of
bulges.




\begin{thebibliography}{}
\bibitem[Bertola et al. 1991]{bertola91} 
     {Bertola, F., Vietri, M., \& Zeilinger, W.~W.} 1991, ApJ, 374, L13 

\bibitem[Fathi \& Peletier 2003]{fathi03} 
     {Fathi, K., \& Peletier, R.~F.} 2003, \textit{A\&A}, 407, 61 

\bibitem[M\'endez-Abreu et al. 2008]{mendezabreu07} {M\'endez-Abreu,
     J., Aguerri, J. A. L., Corsini, E. M. \& Simonneau, E.} 2008,
     A\&A, 478, 353 
     

\end{thebibliography}
\end{document}